\documentclass[journal]{IEEEtran}  

\usepackage{amsfonts,amssymb}  
\usepackage[cmex10]{amsmath}
\usepackage[pdftex]{graphicx}
\usepackage{array,cite,url}

\newtheorem{thm}{Theorem}[section]
\newtheorem{lm}[thm]{\bf Lemma}
\newtheorem{prob}[thm]{\bf Problem}
\newtheorem{df}[thm]{\bf Definition}

\title{Reachable Set Approach to Collision Avoidance for UAVs
}

\author{Yuchen Zhou	and John S. Baras,~\IEEEmembership{Fellow, IEEE}
\thanks{The authors are with the Department of Electrical and Computer Engineering, and the Institute for Systems Research, University of Maryland, College Park, Maryland, USA.%
        email: {\tt\small \{yzh89, baras\}@umd.edu}}
\thanks{Manuscript received Jan 20, 2015}}

\begin{document}

\maketitle
\thispagestyle{empty}
\pagestyle{empty}

\begin{abstract}
In this paper, we propose a reachable set based collision avoidance algorithm for unmanned aerial vehicles (UAVs). UAVs have been deployed for agriculture research and management, surveillance and sensor coverage for threat detection and disaster search and rescue operations. It is essential for the aircraft to have on-board collision avoidance capability to guarantee safety. 
Instead of the traditional approach of collision avoidance between trajectories, we propose a collision avoidance scheme based on reachable sets and tubes. We then formulate the problem as a convex optimization problem seeking suitable control constraint sets for participating aircraft. We have applied the approach on a case study of two quadrotors and two fix-wing aircraft collision avoidance scenario.
\end{abstract}

\section{Introduction}
Autonomous aircraft have been deployed for agriculture research and management, surveillance and sensor coverage for threat detection and disaster search and rescue operations. In most of these scenarios, it is desirable to have multiple aircraft to increase the efficiency and coverage of the UAVs. Since the UAVs in these scenarios, and increasingly in more commercial applications, will be deployed in the shared commercial airspace, they are required to have sophisticated collision avoidance algorithms in order to fly together with other conventional aircraft. As the number of these UAVs increases, a centralized ground control based model is not sufficient alone. Thus an autonomous on-board collision avoidance system needs to be implemented in a decentralized manner.
Many collision avoidance algorithms have been proposed in robotics areas. An artificial potential function was proposed in \cite{leonard_virtual_2001,koren_potential_1991,sigurd_uav_2003} to produce control policies for robots to navigate towards a goal and avoid each other and obstacles. In \cite{carbone_novel_2006,lalish_decentralized_2008,han_proportional_2004}, the authors proposed a decentralized collision avoidance rule based on heading and collision cones. However, the research works mentioned above focus on only designing a single path or trajectory for individual aircraft, so that they are separated by at least the threshold distance. The important challenge remaining is to provide guarantees for unknown execution of the other aircraft, so that aircraft can avoid each other under all possible controls or disturbances in the collaborative setting. An interesting question is whether agents that realize being on a collision path can compute constrained control sets for themselves so that it is guaranteed that all aircraft will not collide under these control constraints and bounded disturbances. This problem implies that the collision avoidance problem needs to solved between time varying sets instead of trajectories.

The reachable set of a dynamical system is defined as the set of states reachable from a given bounded initial set, control set and disturbance set. The practical problem mentioned earlier is closely related to reachability analysis. The collision avoidance problem between reachable sets has been previously studied under the frame of reachability analysis of nonlinear dynamical games \cite{mitchell_time-dependent_2005,vitus_hierarchical_2008}. The other agent is considered as adversary or disturbance to the collision avoidance problem. A controller is synthesized to allow the aircraft to avoid the reachable sets of others. However, in most collision avoidance scenarios, the controller selection is collaborative. In this paper, we investigate cases when agents can derive collaboratively the constrained control sets so that the resulting reachable sets are collision free. 
Besides the above mentioned level set approach \cite{mitchell_time-dependent_2005,vitus_hierarchical_2008} to obtain reachable sets for nonlinear dynamics, there are several fast linear algorithms based on convex analysis. These algorithms employ linearized dynamics with convex initial state set and control disturbance set. They commonly approximate reachable set using specific covering sets including ellipsoids \cite{kurzhanski_ellipsoidal_2000,botchkarev_verification_2000} and polytopes \cite{althoff_reachability_2008,frehse_spaceex:_2011,girard_efficient_2006,asarin_approximate_2000}. In all these cases, commonly support functions are used to analytically derive the reachable sets. However, many algorithms \cite{althoff_reachability_2008,frehse_spaceex:_2011,girard_efficient_2006,asarin_approximate_2000} compute the approximated convex set iteratively, which makes the solutions impossible to be represented in analytic forms. We will use the reachable set tool set from \cite{Kurzhanskiy:EECS-2006-46} based on the ellipsoid methods in \cite{kurzhanski_ellipsoidal_2000}, because its solutions can be expressed efficiently in analytical expressions.

The main contribution of this paper is that we propose and solve a new reachability based formulation of collision avoidance. It is formulated as the following two-fold optimization problem. In the first part, autonomous aircraft collaboratively define control constraint sets, while in the second, individual aircraft will compute an optimal control policy within the control limit so that they can reach their objective and avoid the others. We focus on the first part, since the second part is a traditional optimal control problem with hard control constraints. The rest of the paper is organized as follows. In section \ref{sec:pre} we present the fundamentals of reachability sets. Then in section \ref{sec:prob} we define the reachable set collision avoidance problem and formulate it into a convex optimization problem incorporating the reachable sets. Afterward, we demonstrate our approach in scenarios involving collision avoidance between two quadrotors. 

\section{Preliminaries} \label{sec:pre}
We consider collision avoidance navigation between aircraft whose dynamics are given by nonlinear models as (\ref{eqn1}).
  \begin{equation} \label{eqn1}
   \dot{x}(t)=f(t,x,u,v)
  \end{equation}
where  $x(t) \in \mathcal{X}$, $x(0) \in \mathcal{X}_0 \subseteq \mathcal{X}$, $u(t) \in \mathcal{U}(t)$ for all $t$, $v(t) \in \mathcal{V}$ for all $t$. 

The reachable set of (\ref{eqn1}) (or forward reachable set) $\mathcal{R}[\vartheta]=\mathcal{R}(\vartheta,X_0)$, is the set of states that are reachable at time $\vartheta$ from a set of initial states $X_0$ and all possible controls and disturbances. Formally it is defined by the following,

\begin{df}[Reachable Set] 
The reachable set $\mathcal{R}[\vartheta]=\mathcal{R}(\vartheta,t_0,X_0)$ of the system of (\ref{eqn1}) at time $\vartheta$ from a set of initial positions $X_0$ and time $t_0$ is the set of all points $x$ for which there exists a trajectory $x(s,t_0,x_0), x_0 \in X_0$ that transfers the system from $(t_0, x_0)$ to $(\vartheta,x), x=x(\vartheta)$, while satisfying the associated constraints.
\end{df}

Similarly the reachable tube is the union of all reachable sets over a time interval.
\begin{df}[Reachable Tube]
The reachable tube $\mathcal{R}[\Theta] = \cup_{\vartheta \in\Theta} \mathcal{R}(\vartheta,t_0,X_0)$
\end{df}

Reachable set computation for nonlinear model exists, but either it is impractical in collision avoidance due to slow computation time \cite{mitchell_time-dependent_2005}, or it relies on numerical methods to approximate the nonlinear model with linear models \cite{althoff_reachability_2008}. So instead of looking at the full nonlinear model, we linearize the dynamics around an operating point, resulting in dynamics that are of the following form.
  \begin{equation} \label{eq:linearDyn}
   \dot{x}(t)=A(t)x(t)+B(t)u(t)+v(t)
  \end{equation}
To simplify the computation, the following assumptions are used by noting Lemma \ref{convex}, where $\mathcal{U}(t), \mathcal{X}_0$ and $\mathcal{V}$ are all convex and compact sets. \cite{kurzhanski_ellipsoidal_2000}

\begin{lm}\label{convex}
With $\mathcal{U}(t), \mathcal{X}_0$ and $\mathcal{V}$ being convex and compact, the reachable set $\mathcal{R}[\vartheta]$ is also convex and compact.
\end{lm}

The problem of defining the reachable set of the system can be reformulated as an optimization problem. Consider the system (\ref{eq:linearDyn}). Since the reachable set would be convex and compact due to the assumption on the control set and the initial set, the reachable set can be captured using its support function. Let $\rho(l | X)$ be the support function of the set $X$, i.e. $\rho(l | X)= \max \{\left<l,x\right> | x\in X\}$, and $\left<l,x\right>$ represents the inner product between vector $l$ and $x$. Then the support function of the reachable set $\mathcal{R}[\vartheta]$ is given by the following,
\begin{align}
\rho(l | \mathcal{R}[\vartheta]) &= \max \{\left<l,x\right> | x\in \mathcal{R}[\vartheta]\} \nonumber \\
&= \max\bigg\{ \int_{t_0}^\vartheta {l' \Phi(\vartheta,s)B(s)u(s) + l' \Phi(\vartheta,s)v(s)} ds  \nonumber \\&  \quad+ l' \Phi(\vartheta,t_0)x_0 \bigg| u(s)\in \mathcal{U}(s), x_0\in X_0, v(s) \in \mathcal{V} \bigg\} \nonumber \\
&= \int_{t_0}^\vartheta{ \rho( B'(s)\Phi'(\vartheta,s)l | \mathcal{U}(s))}ds \nonumber \\ 
& \quad \int_{t_0}^\vartheta{ \rho( \Phi'(\vartheta,s)l | \mathcal{V})}ds +\rho(\Phi'(\vartheta,t_0)l | X_0)
\end{align}
where $\Phi(t,s)$ is the transition matrix of the system (\ref{eq:linearDyn}). i.e. it satisfies $\frac{\partial}{\partial t}\Phi(t,s)= A(t)\Phi(t,s)$ and $\Phi(s,s)= \mathbf{I}$. Assume further that all the sets are represented by ellipsoids. Let $c_X$ and $M_X$ denote the center and shape matrix of the set. The following holds, if $x\in X=E(c_X,M_X)$,
\[\left<x-c_X, M_X^{-1}(x-c_X) \right>\leq 1. \]
In terms of the support function, it can be expressed by 
\[\left<l,x\right>\leq \left<l,c_X\right>+\left<l,M_Xl\right>^{1/2}.\]
The support function of the reachable set $\mathcal{R}[\vartheta]$ could be expressed further in terms of the centers and shape matrices of the initial set, control and disturbance sets (equation (\ref{eq:reachableSet})). 
\begin{align} \label{eq:reachableSet}
\rho(l|\mathcal{R}[\vartheta])&= \left<l,\Phi(\vartheta,0)c_{X_0}\right> + \left<l, \int_{0}^{\vartheta}\Phi(\vartheta,s)B(s)c_{\mathcal{U}}(s) ds \right> \nonumber \\& \quad + \left<l,\Phi(\vartheta,0)M_{X_0}\Phi^T(\vartheta,0)l \right>^{1/2} \nonumber \\
&\quad +\left<l, \int_{0}^{\vartheta}\Phi(\vartheta,s)c_{\mathcal{V}}(s) ds \right>\\ \nonumber 
&\quad+ \int_0^\vartheta \left< l, \Phi(\vartheta,s)B(s)M_{\mathcal{U}}B^T(s)\Phi^T(\vartheta,s)l \right>^{1/2} ds \\ \nonumber
&\quad+ \int_0^\vartheta \left< l, \Phi(\vartheta,s)M_{\mathcal{V}}\Phi^T(\vartheta,s)l \right>^{1/2} ds
\end{align}

Since, as will be discussed later, the disturbance is a constant term to the optimization problem, we will assume in what follows that $v(t)=0$. The term related to $v(t)$ affects the size of the reachable set. Since it is independent of the control set parameter, it can be treated as an additional separation required in the collision avoidance problem. 

The computation of reachable set for nonlinear dynamics is similar. Assume the nonlinear system is linearized around a steady operating trajectory $x^*(t)$, with steady state controller $u^*(t)$. Denote total state $X(t)=x^*(t)+ x(t)$. The deviation term $x(t)$ is the linearized dynamics, with state space representation $[A(t),B(t)]$ derived from \ref{eqn1},
\[
A(t) = \left.\frac{\partial f}{\partial x}\right|_{x = x^*}, \quad B(t) = \left. \frac{\partial f}{\partial u} \right|_{u=u^*}
\]
The reachable set of the original nonlinear system have the same shape but with additional center offset based on the steady state dynamics $x^*(t)$ which evolves according to the following.
\[
\dot{x}^*(t) = f(t, x^*, u^*) 
\]
In the following section, we assume the system is a linear dynamics. In the simulation section, we provide a nonlinear example. The method is very similar but with few constraints adjusted accordingly due to the steady state dynamics.

\section{Collision Avoidance Between Two Agents Using Reachability Analysis}\label{sec:prob}

Let us consider the following two agents reachability based collision avoidance problem.

\begin{prob}[Reachability Based Collision Avoidance] 
We seek a control algorithm for the aircraft A and B, such that they can always avoid each other if each of them uses a more constrained control set than their initial ones. More specifically, in the first phase, given the initial state estimation set of the aircraft B $X^B_0$, and the estimated control set $\mathcal{U}^B$, we seek a tighter control constraint set $\tilde{\mathcal{U}}^B$ such that aircraft A can find a safe reachable tube that does not intersect with the one of aircraft B counting the separation. At the same time, we need to make sure that under the new control constraint set $\tilde{\mathcal{U}}^B$, aircraft B can still perform the maneuvers the system requires to reach its goal area. At the second phase we seek a safe reachable tube for aircraft A so that the reachable tube will be apart from the reachable tube of aircraft B for at least the required separation. Afterward we seek trajectories within the reachable tube, so that they can safely reach their objectives in an optimal manner. 
\end{prob}

As assumed in the previous section, let the initial set and control set of aircraft B be the ellipsoids $X^B_0=E(c_{X_0}^B,M_{X_0}^B)$, $\mathcal{U}^B = E(c_{U}^B,M_{U}^B)$, and denote by $c_{X}^A(t)$ the nominal trajectory of aircraft A. We further assume the two aircraft will be closest at time $\tau$ in the future, so we can reason about the reachable set instead of the tube. For the first phase we want to keep the reachable set of aircraft B away from the nominal trajectory of A as far as we can, and keep the constrained control set of aircraft B relatively large. This can be formulated into the following optimization problem.

\begin{prob}[Optimization Part I] \label{prob:controlSet}
\[\begin{array}{c c} \underset{q(t),Q(t)}{\max} \underset{l}{\max}  \text{ }   & \left<l,c_{X}^A(\tau)\right>-\rho(l | \mathcal{R}^B_{x}(\tau)) + k \log (\det (Q(t))) \\
 \text{subject to  } & E(q(t),Q(t)^TQ(t)) \subset \mathcal{U}^B \quad \forall t \in [0,\tau] \\
 &\left\|l\right\|=1 \\
& \left<l,c_{X}^A(\tau)\right>-\rho(l | \mathcal{R}^B_{x}(\tau)) \geq 0
 \end{array}\]
\end{prob}
where $\rho(l|\mathcal{R}_{x}^B(\tau))$ is the support function of the reachable set of aircraft B at time $\tau$:
\begin{align} \label{eq:reachableSetB}
\rho(l|&\mathcal{R}^B_{x}(t))= \left<l,\Phi(t,0)c^B_{X_0}\right> + \left<l, \int_{0}^{t}\Phi(t,s)B(s)c_U^B(s) ds \right> \nonumber \\& \quad + \left<l,\Phi(t,0)M^B_{X_0}\Phi^T(t,0)l \right>^{1/2} \\ \nonumber 
&+ \int_0^t \left< l, \Phi(t,s)B(s)Q^T(s)Q(s)B^T(s)\Phi^T(t,s)l \right>^{1/2} ds.
\end{align}

The parameters of the optimization are $q(t)$ and $Q(t)$, both related to the new control set. More specifically $\tilde{\mathcal{U}}^B=E(q(t),Q^T(t)Q(t))$. In the objective function, the first part is the distance between the nominal trajectory $c_{X}^A(\tau)$ and the reachable set, and the second part is the size of the control set. Scalarization is used to determine the Pareto optimal points \cite{boyd2004convex}. The inner maximization determines distance by varying the direction vector $l$. The first constraint is due to the fact that $\tilde{\mathcal{U}}^B \subset \mathcal{U}^B$, the last constraint is to keep nominal trajectory outside the reachable set of aircraft B. 

$ E(q(t),Q(t)^TQ(t)) \subset \mathcal{U}^B$ is equivalent to the following constraints on a new parameter $\lambda >0$,
\[\begin{bmatrix} 
   1-\lambda & 0 & (q(t)-c^B_{U})^T \\
	 0 & \lambda I & Q(t) \\
	q(t)-c^B_{U} & Q(t) & M^B_{U}
	\end{bmatrix} \succeq 0.
\]

Let $A$, $B$ be time invariant, we seek a time invariant control constraint set as well, so $q$ and $Q$ are time invariant.
Let us assume the optimal $l^*$ can be estimated based on the initial center of the reachable set alone. In other words, we assume the direction that minimizes distance between the reachable set and $c_X^A(\tau)$ is not affected by changes in the control constraint. The intuition behind this assumption is that even if the direction is altered, the outer maximization is achieved at a similar constraint set. In real applications, the autonomous aircraft will be given such direction to avoid either based on the approaching angle autonomously or based on instructions from the other pilots.
Then the overall problem becomes the following.
\begin{prob}[Simplified Problem Part I]
\[\begin{array}{c c} \underset{q,Q}{\max} \text{ }   & \left<l^*,c_{X}^A(\tau)\right>-\rho(l^* | \mathcal{R}^B_{x}(\tau)) + k(\log \det (Q)) \\
 \text{subject to  } & \lambda > 0 \\
& \begin{bmatrix} 
   1-\lambda & 0 & (q-c^B_{U})^T \\
	 0 & \lambda I & Q \\
	q-c^B_{U} & Q & M^B_{U}
	\end{bmatrix} \succeq 0 \\
& \left<l^*,c_{X}^A(\tau)\right>-\rho(l^* | \mathcal{R}^B_{x}(\tau)) \geq 0
 \end{array}\]
\end{prob}
\begin{align}
\rho(l|\mathcal{R}^B_{x}(t))&= \left<l,e^{At} c^B_{X_0}\right> + \left<l, \int_{0}^{t}e^{A(t-s)}ds Bq \right>\nonumber \\
&+ \left<l,e^{At}M^B_{X_0}e^{At}l \right>^{1/2} \\ \nonumber 
&+  \int_0^t \left< l, e^{A(t-s)} B Q^T Q B^T (e^{A(t-s)})^T l \right>^{1/2} ds.
\end{align}

To make the overall problem a simple convex optimization problem, we note the following two cases. 

(i) Suppose $Q$ is $r* (M^B_{U})^{1/2}$ i.e. the constrained control set is a scaled version of the initial control set. Then we have the following.
	\begin{prob}[Scaled Initial Set Method]
\[\begin{array}{c c} \underset{q,r}{\max} \text{ }   & \left<l^*,c_{X}^A(\tau)\right>-\rho(l^* | \mathcal{R}^B_{x}(\tau)) \\
&+ k \, \text{dim}(M^B_{U}) \log r \\
 \text{subject to  } & \lambda > 0 \\
& \begin{bmatrix} 
   1-\lambda & 0 & (q-c^B_{U})^T \\
	 0 & \lambda I & r(M^B_{U})^{1/2} \\
	q-c^B_{U} & r(M^B_{U})^{1/2} & M^B_{U}
	\end{bmatrix} \succeq 0 \\
& \left<l^*,c_{X}^A(\tau)\right>-\rho(l^* | \mathcal{R}_{xB}(\tau)) \geq 0
 \end{array}\]
\end{prob}
\begin{align}
\rho(l|\mathcal{R}^B_{x}&(t))= \left<l,e^{At} c^B_{X_0}\right> + \left<l, \int_{0}^{t}e^{A(t-s)}ds Bq \right>\nonumber \\
&+ \left<l,e^{At} M^B_{X_0} e^{At}l \right>^{1/2} \\ \nonumber 
&+  r\int_0^t \left< l, e^{A(t-s)} B M^B_{U} B^T (e^{A(t-s)})^T l \right>^{1/2} ds.
\end{align}

(ii) Assume now that the requirement for $\left<l^*,c_{X}^A(\tau)\right>-\rho(l^* | \mathcal{R}^B_{x}(\tau)) \geq 0$ is removed. The last term of $\rho(l^* | \mathcal{R}^B_{x}(\tau))$ can be upper bounded using the matrix norm property. Then the objective function can be lower bounded by $\left<l^*,x_{Ac}\right> -\tilde{\rho}(l^* | \mathcal{R}^B_{x}(\tau))$, where
\begin{align}
\tilde{\rho}(l|\mathcal{R}^B_{x}&(t))= \left<l,e^{At} c^B_{X_0}\right> + \left<l, \int_{0}^{t}e^{A(t-s)}ds Bq \right>\nonumber \\
&+ \left<l,e^{At} M^B_{X_0} e^{At}l \right>^{1/2} \\ \nonumber 
&+  \left\|Q\right\|_2 \int_0^t \left< l, e^{A(t-s)} B B^T (e^{A(t-s)})^T l \right>^{1/2} ds.
\end{align}
As the result, we have the following convex optimization problem. 
	\begin{prob}[Matrix Norm Method]
\[\begin{array}{c c} \underset{q,Q}{\max} \text{ }   & \left<l^*,c_{X}^A(\tau)\right>-\tilde{\rho}(l^* | \mathcal{R}^B_{x}(\tau))\\ &+ k \log(\det(Q)) \\
 \text{subject to  } & \lambda > 0 \\
& \begin{bmatrix} 
   1-\lambda & 0 & (q-c^B_{U})^T \\
	 0 & \lambda I & Q \\
	q-c^B_{U} & Q & M^B_{U}
	\end{bmatrix} \succeq 0 \\
& \left<l^*,c_{X}^A(\tau)\right>-\tilde{\rho}(l^* | \mathcal{R}^B_{x}(\tau)) \geq 0
 \end{array}\]
\end{prob}

Both methods can be solved easily by a convex optimization solver in particular a semidefinite programming solver. We use CVX \cite{cvx} in the demonstration discussed in a later section. After the control set of aircraft B is determined, aircraft A can define its control set so that the reachable set does not collide with the safe set of B. The safe set of B is the reachable set of B enlarged by the required separation between aircraft A and B. The difference with the previous problem is the fact that we cannot use the center of aircraft A to maximize the distance to the reachable set of B. Instead we add a constraint to keep the distance between the sets larger than zero. If such a problem is feasible, standard optimization algorithms will be used to obtain control laws for both aircraft under the modified constrained control set. If it is not feasible, then it means the scalarization factor is too large. By iteratively decreasing the scalarization factor, one can find a pair of good control sets that keep the reachable set of both aircraft balanced. After they avoid each other on the closest contact point, the original navigation will resume. 

The optimization part II is addressing the problem to obtain the largest control set for aircraft A. After the constrained control set of aircraft B is determined, the external ellipsoid approximation of the reachable set of aircraft B will be computed based on the ellipsoid toolbox \cite{Kurzhanskiy:EECS-2006-46}.  The external approximation is given as an intersection of ellipsoids with the same center. The safe set of B is defined as the Minkowski sum of the reachable set of aircraft B and a ball, whose radius is the required separation. It can be computed by the ellipsoid toolbox as the intersection of overapproximated ellipsoids as well. For simplicity, we take an external ellipsoid approximation of the intersection set as the safe set of B. Let such overapproximation ellipsoid be $E(c_{X}^B,M_{X}^B)$. Assume that the direction for which the minimum distance is achieved is $l^*$ as assumed earlier. Then the largest control set for aircraft A that satisfies the safety requirement can be determined by the following optimization problem.
\begin{prob}[Optimization Part II] \label{prob:controlSetB}
\[\begin{array}{c c} \underset{q,Q}{\max}  \text{ }   &  \log (\det (Q)) \\
 \text{subject to  } & E(q,Q^TQ) \subset \mathcal{U}^A \\
&-\rho(-l^*|\mathcal{R}^A_{x}(\tau))-\rho(l^*|E(c_{X}^B,M_{X}^B))>0 
 \end{array}\]
\end{prob}
This can be again transformed into a convex optimization problem by using the shrinking initial set or the norm method. We will focus on the norm method for this part. The last constraint can be reformulated into the following:
\begin{multline*}
\left<l^*,e^{A\tau} c^A_{X_0}\right> + \left<l^*, \int_{0}^{\tau}e^{A(\tau-s)}ds Bq \right>-\left<l^*,c_{X}^B\right>\\
- \left<l^*,e^{A\tau} M^A_{X_0} e^{A\tau}l^* \right>^{1/2}  - \left<l^*,M_{X}^Bl^*\right>^{1/2}\\
-  \left\|Q\right\|_2 \int_0^t \left< l^*, e^{A(\tau-s)} B B^T (e^{A(\tau-s)})^T l^* \right>^{1/2} ds  >0.
\end{multline*}

\section{Simulations and Results}
The reachable set based method described above is demonstrated on the linearized quadrotor models described below.
\subsection{Quadrotor Model}
To capture the dynamics of the quadrotor properly, we need two coordinate frames. One of them is a fixed frame and will be named as the earth frame, and the second one is the body frame which moves with the quadrotor. The transformation matrix from the body frame to the earth frame is $R(t)$. 
The quadrotor dynamics has twelve state variables ($x,y,z,v_x,v_y,v_z,\phi,\theta,\psi,p,q,r$), where $\xi=[x,y,z]^T$ and $v=[v_x,v_y,v_z]^T$ represent the position and velocity of the 
quadrotor w.r.t the body frame. $(\phi,\theta,\psi)$ are the roll, pitch and yaw angles, and $\Omega=[p,q,r]^T$ are the rates of change of roll, pitch and yaw respectively.  

The Newton-Euler formalism for the quadrotor rigid body dynamics in earth fixed frame is given by: 
\begin{align} \label{quadrotor}
 \dot{\xi} & =v  \nonumber \\
 \dot{v} & = -g\mathbf{e_3} + \frac{F}{m}R\mathbf{e_3}\\ \nonumber
 \dot R &=R \hat{\Omega} \\ \nonumber
 \dot\Omega &= J^{-1} (-\Omega \times J\Omega + u)
\end{align}
where $g$ is the acceleration due to gravity, $\mathbf{e_3}=[0, 0, 1]^T$, $F$ is the total lift force and $u=[u_1,u_2,u_3]^T$ are the torques applied. $F$ and $u$ are the control inputs. $J$ is the moment of inertial details on the quadrotor dynamics can be
found in \cite{Kumar},\cite{Garcia}.
For this work, we linearize the dynamics (\ref{quadrotor}) about the hover with yaw constraint to be zero, as it has been done in \cite{WolffICRA}. Since $\psi$ is constrained to be zero, 
we remove $\psi$ and $r$ from our system and make the system ten dimensional. Consequently, we only need three control inputs, $F, u_1$, and $u_2$ for the system. The linearized model is the same as what is done in  \cite{WolffICRA}, \cite{Dustin}. The system matrices for the linearized
 model are: \\
 $A=\begin{bmatrix} \mathbf{0} & I & \mathbf{0} & \mathbf{0} \\ 
 \mathbf{0} & \mathbf{0} & \begin{bmatrix} 0  & g \\ -g & 0\\0 & 0 \end{bmatrix} & \mathbf{0}\\ 
 \mathbf{0} &\mathbf{0} &\mathbf{0} & I \\ \mathbf{0} &\mathbf{0} &\mathbf{0} &\mathbf{0}\\
 \end{bmatrix}; ~ B = \begin{bmatrix} \mathbf{0} & \mathbf{0}\\ \begin{bmatrix} 0\\  0 \\ 1/m \end{bmatrix} &\mathbf{0} \\ \mathbf{0} & \mathbf{0} \\\mathbf{0} & I_{2 \times 3} J^{-1}\\ \end{bmatrix}$
 $I_{2,3}= \begin{bmatrix}  1 & 0 &0\\0 &1 &0 \\\end{bmatrix}$ \\
 All zero and identity matrices in $A$ and $B$ are of proper dimensions.  

\subsection{Fix-wing Dynamics Model}
Similar to the dynamics of the quadrotor, we used the earth frame and the body frame. Let $(\phi,\theta,\psi)$ be roll pitch yaw similar as before and $(p,q,r)$ be the body frame rotation rates. The transformation matrix from the body frame to the earth frame is $R$. The velocity state is commonly represented in the body frame as $v_b = (u,v,w)$.
\begin{align} \label{fixedwing}
 \dot{\xi} & =R v_b  \nonumber \\
 \dot{v}_b & = 
\begin{pmatrix}
rv-qw \\
qw -ru \\
qu- pv
\end{pmatrix} + \frac{1}{m}\begin{pmatrix}
F_x \\
F_y \\
F_z
\end{pmatrix}  \\
 \dot R &=R \hat{\Omega} \nonumber \\ \nonumber
 \dot\Omega &= J^{-1} (-\Omega \times J\Omega + T)
\end{align}
where $(F_x, F_y, F_z)$ is the force in body frame. $T$ is the torque in body frame. $J$ is the moment of inertial. The force and torque is further related to airspeed and actuator. Due to the decoupling between lateral and longitude motion in conventionally aircraft, we will focuses on longitude motion here. The control signal for longitude motion are elevators control $\delta_e$ and propeller thrust $\delta_t$. We can obtain the following linearized dynamics around the leveled cruise mode $x^*$ and $u^*$. In the leveled cruise mode, there is no turning or climbing. The longitude motion can be captured in a 6 state system ($x,z,u,w,q,\theta$) and 2 control ($\delta_e, \delta_t$). The linearized dynamics can be represented in the matrix form.

\[A=\begin{bmatrix} 
0 & 0 & 1 & 0 & 0 & 0 \\
0 & 0 & \sin{\theta^*} & -\cos{\theta^*} & 0 & u^*\cos{\theta^*} + w^* \sin{\theta^*} \\
0 & 0 & X_u & X_w & X_q & -g \cos{\theta^*} \\
0 & 0 & Z_u & Z_w & Z_q & -g \sin{\theta^*} \\
0 & 0 & M_u & M_w & M_q & 0 \\
0 & 0 & 0 & 0 & 1 & 0 \\
 \end{bmatrix}\]
\[B = 
\begin{bmatrix}
 0 & 0 \\
 0 & 0 \\
 X_{\delta_e} & X_{\delta_t} \\
 Z_{\delta_e} & 0 \\
 M_{\delta_e} & 0 \\
 0 & 0 \\
\end{bmatrix}\]

All the terms of the linear dynamics are derived from aerodyanmic models that can be found in \cite{beard2012small}. 

\subsection{Reachable Sets and Constrained Control Sets}

We computed the reachable sets and performed convex optimization using a computer with a 3.4GHz processor and 8GB memory. The software we used include the Ellipsoid toolbox \cite{Kurzhanskiy:EECS-2006-46} and CVX \cite{cvx,gb08} for solving convex optimization problems.

\subsubsection{Quadrotor Example}
\begin{figure}
	\centering
		\includegraphics[width=0.9\linewidth]{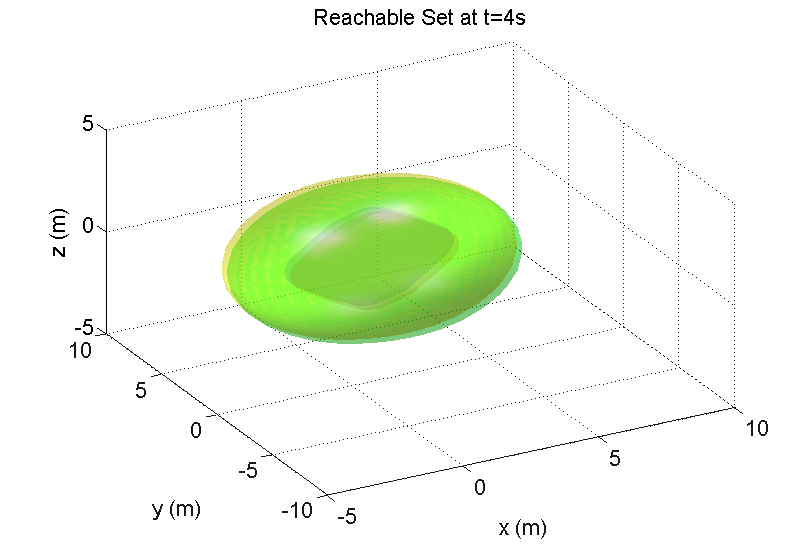}
		\caption{The initial reachable sets of both aircraft projected to $x$ $y$ $z$ axis at time $\tau$. The two reachable sets are represented by the internal and external approximations. The reachable sets largely overlap each other. The light colored sets are the external approximations of the reachable sets of aircraft A and B. The internal approximations are the darker color ones inside.}
	\label{fig:initSet}
\end{figure}

\begin{figure}
	\centering
		\includegraphics[width=0.9\linewidth]{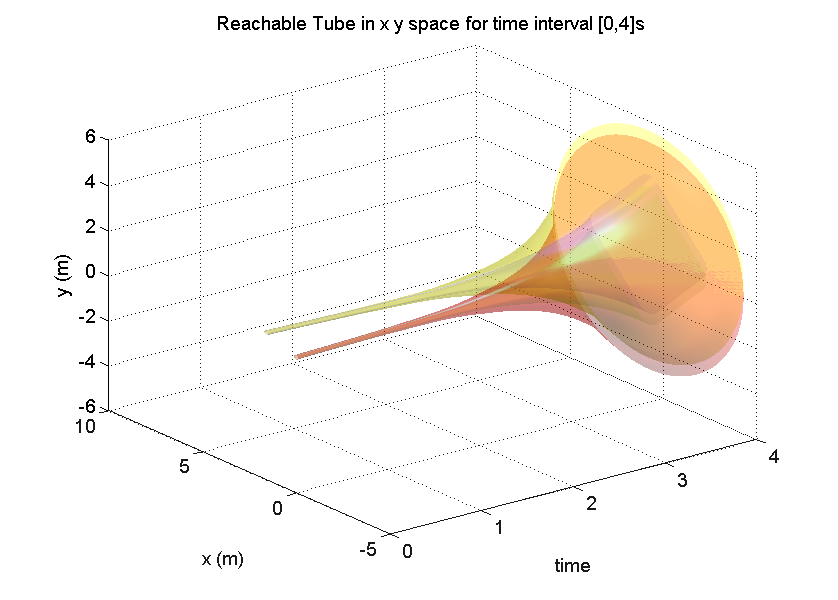}
		\caption{The initial reachable tubes of of both aircraft projected to $x$ $y$. Again the reachable tubes are represented by the internal and external approximations. The light yellow tube is the external approximation of the reachable tube of aircraft A, while the light red one is the external approximation of the reachable tube of aircraft B. The internal approximations are the darker color ones inside. Clearly, the reachable tubes collide.}
	\label{fig:initTube}
\end{figure}

\begin{figure}
	\centering
		\includegraphics[width=0.9\linewidth]{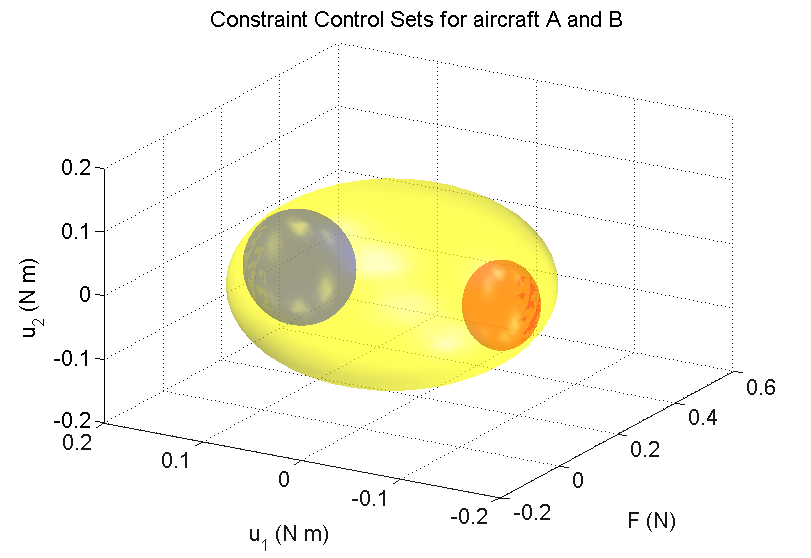}
		\caption{This shows the initial control set (yellow ellipsoid), the constrained control set for aircraft A (red ellipsoid), and the constrained control set for aircraft B (blue ellipsoid). This pair of constrained control sets is obtained when the scalarization factor is 1. The control set for $u_1$, which contributes to yaw rotation, of aircraft B is larger, i.e. this fits the case when aircraft B has higher priority, so it has more freedom in terms of maneuvers.}
	\label{fig:constraintk1}
\end{figure}

\begin{figure}
	\centering
		\includegraphics[width=0.9\linewidth]{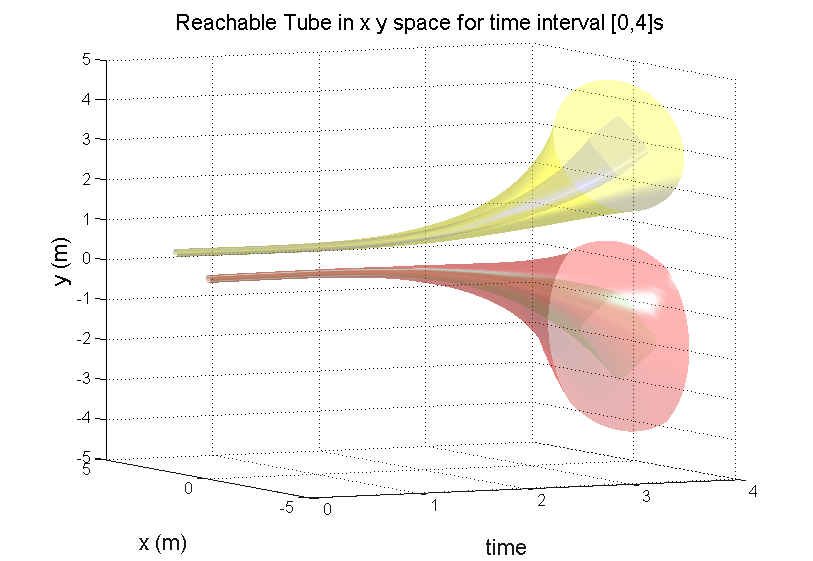}
		\caption{The reachable tubes for aircraft A and B with $k=1$. Clearly, there is no collision between the overapproximation of the reachable tubes. As can be seen from the reachable tube as well, the aircraft A has less freedom comparing to aircraft B. A closer examination also reveals that there is no violation of separation requirement over time.}
	\label{fig:endTubek1}
\end{figure}

\begin{figure}
	\centering
		\includegraphics[width=0.9\linewidth]{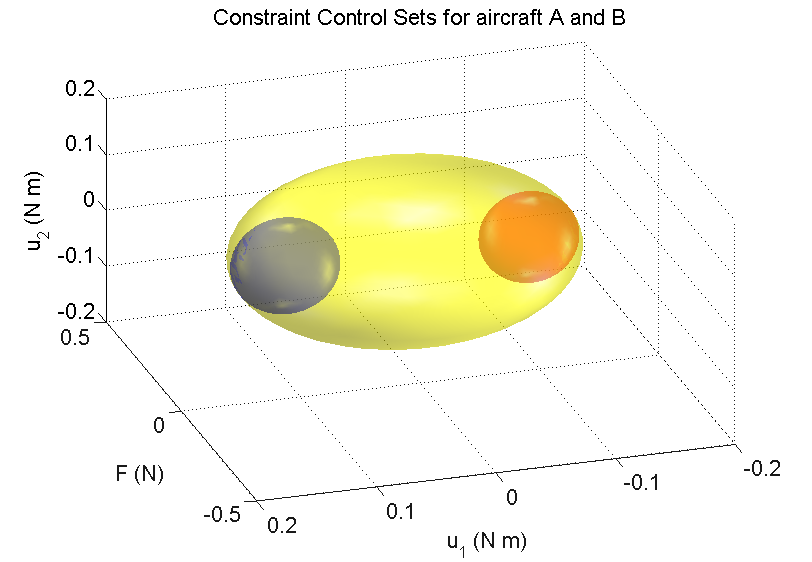}
		\caption{This shows the initial control set (yellow ellipsoid), the constrained control set for aircraft A (red ellipsoid), and the constrained control set for aircraft B (blue ellipsoid). This pair of constrained control sets is obtained when the scalarization factor is 0.9. The control set sizes are rather balanced.}
	\label{fig:constraintk0.9}
\end{figure}

\begin{figure}
	\centering
		\includegraphics[width=0.9\linewidth]{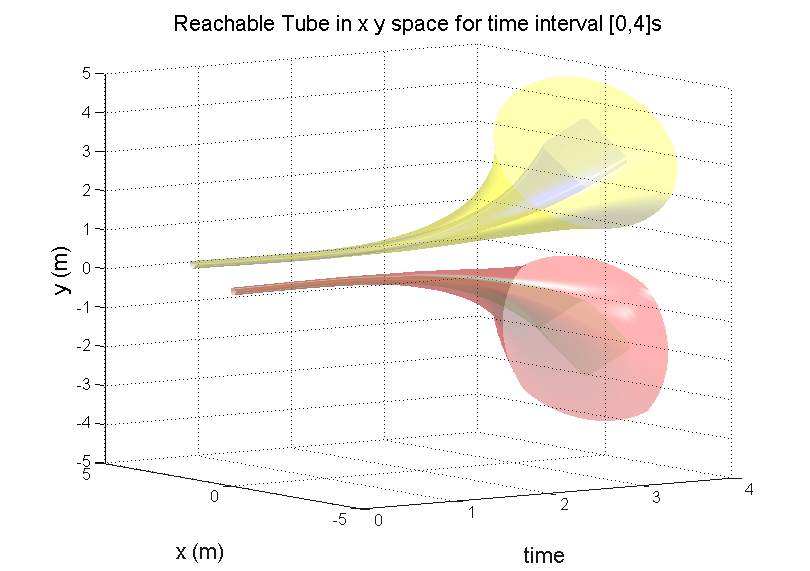}
		\caption{The reachable tubes for aircraft A and B with $k=0.9$. Clearly, there is no collision between the overapproximation of the reachable tubes. A closer examination also reveals that there is no violation of separation requirement over time.}
	\label{fig:endTubek0.9}
\end{figure}

The first scenario is the following: two quadrotors are approaching each other from coordinates around $(1.6, 0.5, 0)$m and $(0,0,0)$m with initial speed $(-0.2,0,0)$m/s, and $(0.2,0,0)$m/s respectively. The separation requirement is $1$m. It is fairly easy to estimate the collision time which is $\tau=4$s, and the closest direction $l^*$. The initial reachable tube of both aircraft projected to the $x$, $y$ and time axis is shown in Fig. \ref{fig:initTube}. The reachable sets at time $\tau$ are shown in Fig. \ref{fig:initSet}. As can be seen, the reachable sets clearly overlap with each other. There are no guarantees that one can obtain from this initial setup. By varying $k$ in the optimization problem \ref{prob:controlSet}, we get the following pairs of control sets for aircraft A and B. (Fig. \ref{fig:constraintk1}, \ref{fig:constraintk0.9}). The control sets are obtained using the matrix norm method. The corresponding resulting reachable tubes for agent A and B are plotted in Fig. \ref{fig:endTubek1}, \ref{fig:endTubek0.9}. Clearly, the resulting reachable tubes avoid each other, and a closer examination shows that the separation requirement is satisfied. 

The computation time for the overall problem is dominated by the reachable set computation. Since we would like to have good precision of the reachable set, 30 ellipsoids are used to obtain the external approximation for both aircraft. The computation for all the reachable sets takes around 430s. In this case study, the collision time is almost imminent (4s). It is only possible to implement this in real time by using a poor estimate of the reachable set with lower precision in this setup. The constrained control sets obtained that way are much smaller. It is also possible to compute the reachable sets in parallel so the resulting computation time would be almost 4s. There are other ways to present the approximation such as via zonotope \cite{althoff_reachability_2008} or support function \cite{frehse_spaceex:_2011}. Preliminary testing using a zonotope based method shows faster computation time, but we have not obtained a good representation of the constrained optimization problem \ref{prob:controlSet} and \ref{prob:controlSetB} using zonotopes. One of our future directions is to use other representations to make the process faster.

However, if one does not require such reachable sets to be computed for real time verification, the computation of the control sets by itself takes very little amount of time. 4s is definitely enough to derive such control sets through convex optimization.  

\subsubsection{Fix-wing Aircraft Example}

\begin{figure}
	\centering
		\includegraphics[width=0.9\linewidth]{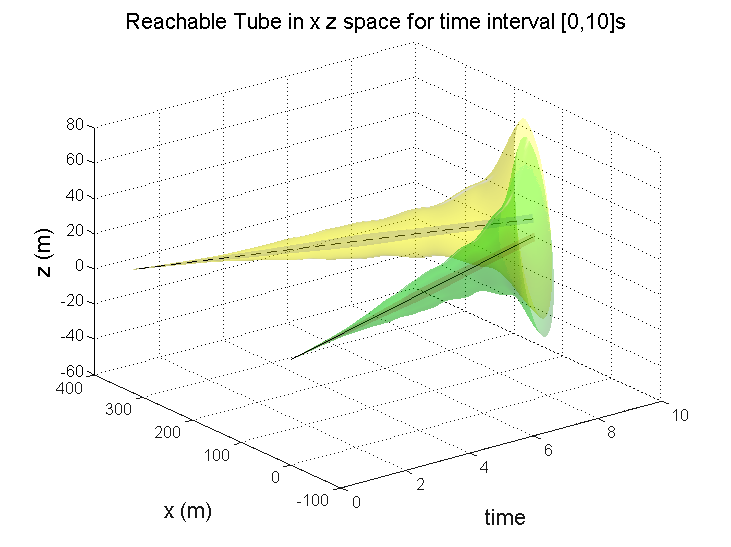}
		\caption{The initial reachable tubes of of both aircraft projected to $x$ $z$. The reachable tubes are represented by the internal and external approximations. The light yellow tube is the external approximation of the reachable tube of aircraft A, while the light green one is the external approximation of the reachable tube of aircraft B. The internal approximations are the darker color ones inside. Clearly, the reachable tubes collide. The steady state trajectory for aircraft A is specified by dashed black line, and the steady state trajectory of aircraft B is specified by the blue line.}
	\label{fig:initTube_fixwing}
\end{figure}

\begin{figure}
	\centering
		\includegraphics[width=0.9\linewidth]{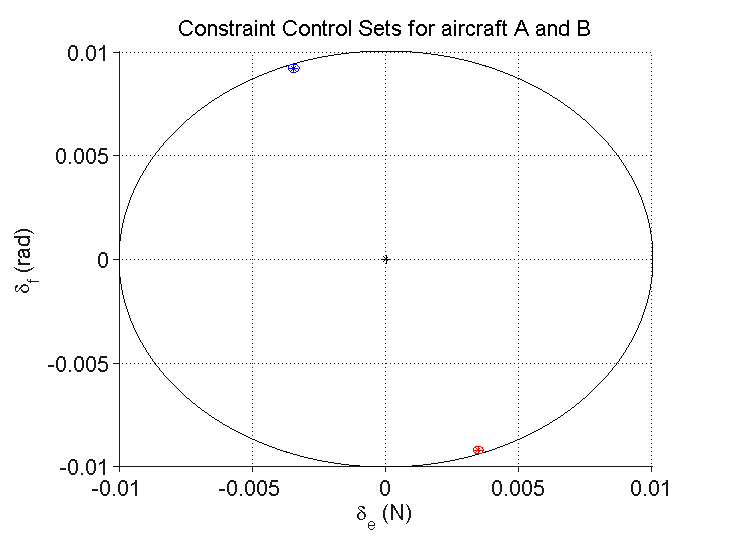}
		\caption{This shows the initial control set (black ellipse), the constrained control set for aircraft A (red ellipse), and the constrained control set for aircraft B (blue ellipse). The ellipses are specified by centers and boundaries with their corresponding colors. The constrained control sets are much smaller comparing to the original one, so only center of the ellipse are very clear.}
	\label{fig:constraint_fixwing}
\end{figure}

\begin{figure}
	\centering
		\includegraphics[width=0.9\linewidth]{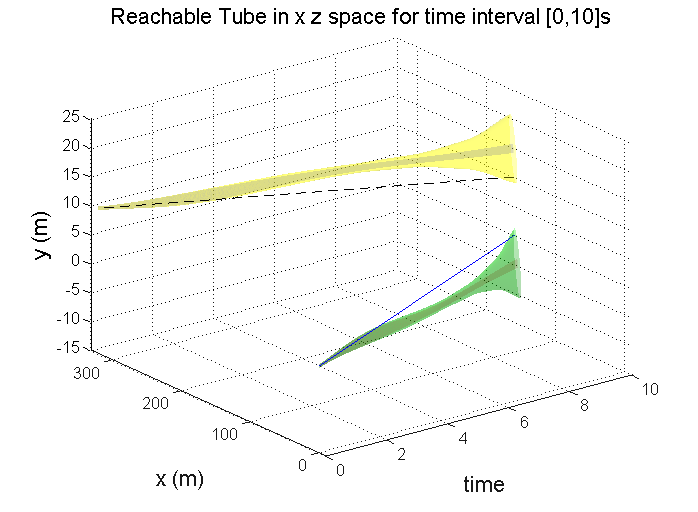}
		\caption{The reachable tubes for aircraft A and B Clearly, there is no collision between the overapproximation of the reachable tubes. A closer examination also reveals that there is no violation of separation requirement over time.}
	\label{fig:endTube_fixwing}
\end{figure}

The second scenario is the following: two fix-wing aircraft are approaching each other from coordinates around $(320, 10, 0)$m and $(0,0,0)$m with initial speed $(-16,0,0)$m/s, and $(16,0,0)$m/s respectively. The separation requirement is $10$m. The collision time is approximately $\tau=10$s, and the closest direction $l^*$ is in the z axis. The initial reachable tube of both aircraft projected to the $x$, $z$ and time axis is shown in Fig. \ref{fig:initTube_fixwing}. As can be seen, the reachable sets clearly overlap with each other. There are no guarantees that one can obtain from this initial setup. By varying $k$ in the optimization problem \ref{prob:controlSet}, we get the following pair of control sets for aircraft A and B. (Fig. \ref{fig:constraint_fixwing}). The control sets are obtained using the matrix norm method. The corresponding resulting reachable tubes for agent A and B are plotted in Fig. \ref{fig:endTube_fixwing}. Clearly, the resulting reachable tubes avoid each other, and a closer examination shows that the separation requirement is satisfied. 

\section{Conclusion}

In this paper, we have proposed a reachability based approach to collision avoidance of UAVs so that the resulting reachable tubes are collision free. Our approach provides new insight to the collision avoidance problem in particular regarding collision avoidance of set of trajectories instead of one. Furthermore, as this approach gives limited constraints on the controller, standard optimization based or rule based controller design can be used after our method to obtain optimal and safe trajectories. Our approach reformulates the collision avoidance problem to a two-fold convex optimization problem, which can be solved very efficiently. The computation time can be dramatically improved by parallel computation and other reachable set analysis tools. 

Although currently, the method is limited to linear systems, development of reachability analysis for hybrid systems can extend our method to nonlinear and more general dynamics. We focused the analysis for two aircraft collision avoidance, but our method can be extended to the multiple aircraft case fairly easily by iteratively fixing the control set and the resulting reachable set one by one. 

\section*{ACKNOWLEDGMENT}
This work is supported by US AFOSR MURI grant FA9550-09-1-0538,
NSF grant CNS-1035655, NIST grant 70NANB11H148 and by DARPA (through ARO) grant W911NF1410384.


\bibliographystyle{IEEEtran}
\bibliography{bib}

\begin{IEEEbiography}
    [{\includegraphics[width=1in,height=1.25in,clip,keepaspectratio]{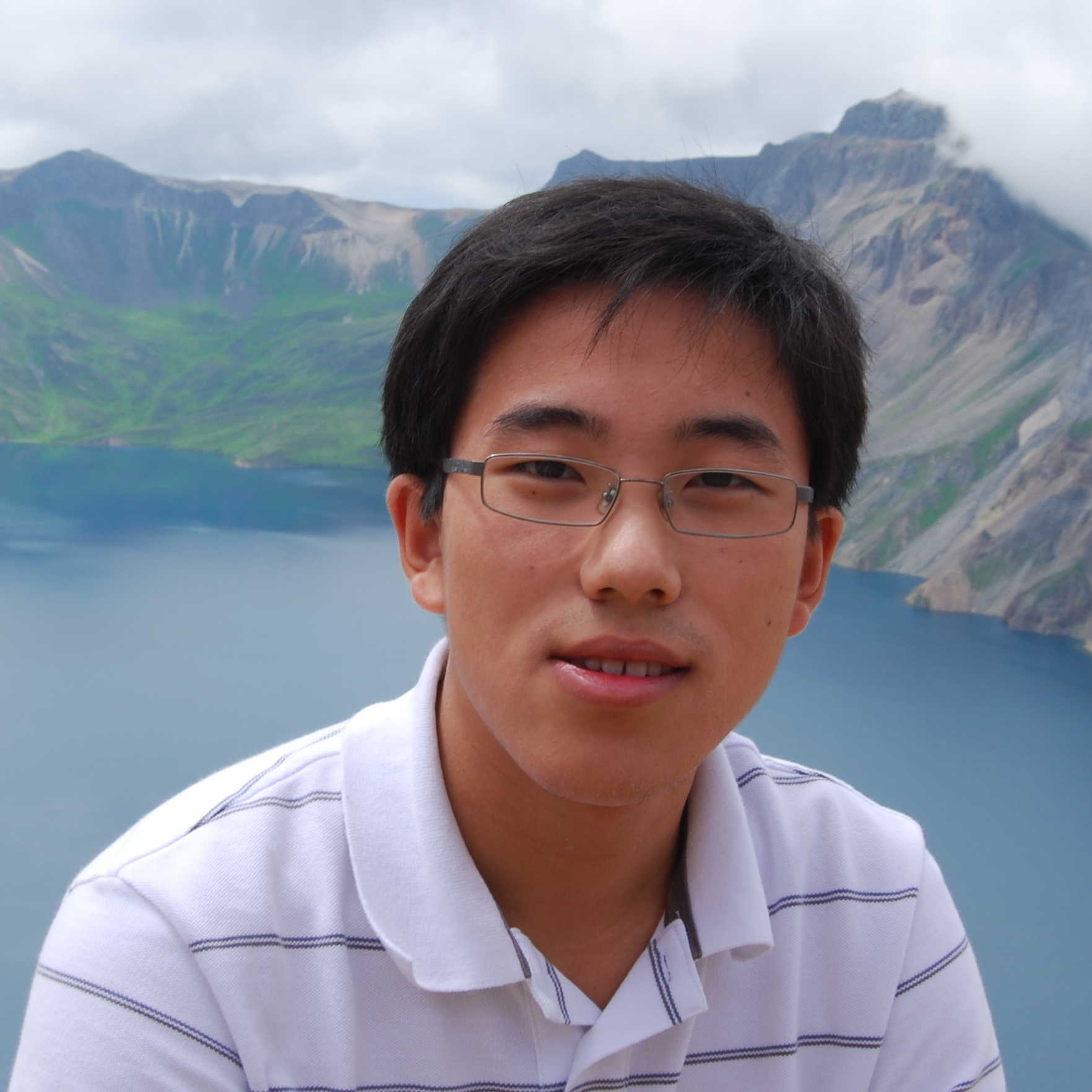}}]{Yuchen Zhou} received the B.S. degree with honors in Electrical Engineering from the University of Maryland, College Park, in 2011 and he received the Ph.D. degree in Controls/Electrical Engineering from the University of Maryland College Park, USA, in July 2016. His research interests include timed temporal logic based planning, robotic motion planning, optimal controls and reachability analysis.  In Spring 2015, he was an exchange PhD student at KTH, Royal Institute of Technology, Stockholm, Sweden.
\end{IEEEbiography}

\begin{IEEEbiography}[{\includegraphics[width=1in,height=1.25in,clip,keepaspectratio]{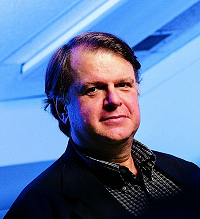}}]{John S. Baras}
received the Diploma in Electrical
and Mechanical Engineering with highest distinction
from the National Technical University of Athens,
Greece, in 1970. He received the M.S. and Ph.D.
degrees in Applied Mathematics from Harvard University,
Cambridge, MA, in 1971 and 1973 respectively.
Since 1973 he has been with the Department
of Electrical and Computer Engineering, University
of Maryland at College Park, where he is currently
Professor, member of the Applied Mathematics,
Statistics and Scientific Computation Program Faculty,
and Affiliate Professor in the Fischell Department of Bioengineering and
the Department of Mechanical Engineering. From 1985 to 1991 he was the
Founding Director of the Institute for Systems Research (ISR) (one of the first
six NSF Engineering Research Centers). In February 1990 he was appointed
to the Lockheed Martin Chair in Systems Engineering. Since 1991 Dr. Baras
has been the Director of the Maryland Center for Hybrid Networks (HYNET),
which he co-founded. Dr. Baras has held visiting research scholar positions
with Stanford, MIT, Harvard, the Institute National de Reserche en Informatique
et en Automatique (INRIA), the University of California at Berkeley,
Linkoping University and the Royal Institute of Technology (KTH) in Sweden.
Among his awards are: the 1980 George S. Axelby Award of the IEEE Control
Systems Society; the 1978, 1983 and 1993 Alan Berman Research Publication
Awards from the NRL; the 1991, 1994 and 2008 Outstanding Invention of
the Year Awards from the University of Maryland; the 1998 Mancur Olson
Research Achievement Award, from the Univ. of Maryland College Park;
the 2002 and 2008 Best Paper Awards at the 23rd and 26th Army Science
Conferences; the 2004 Best Paper Award at the Wireless Security Conference
WISE04; the 2007 IEEE Communications Society Leonard G. Abraham Prize
in the Field of Communication Systems; the 2008 IEEE Globecom Best
Paper Award for wireless networks; the 2009 Maryland Innovator of the
Year Award. In November 2012 he was honored by the Awards for both
the Principal Investigator with Greatest Impact and for the Largest Selling
Product with Hughes Network Systems for HughesNet, over the last 25 years
of operation of the Maryland Industrial Partnerships Program. These awards
recognized Dr. Baras pioneering invention, prototyping, demonstration and
help with commercialization of Internet protocols and services over satellites
in 1994, which created a new industry serving tens of millions worldwide.
In 2014 he was awarded the 2014 Tage Erlander Guest Professorship by the
Swedish Research Council, and a three year (2014-2017) Hans Fischer Senior
Fellowship by the Institute for Advanced Study of the Technical University
of Munich. Dr. Baras has been the initial architect and continuing innovator
of the pioneering MS on Systems Engineering program of the ISR. Dr. Baras
research interests include control, communication and computing systems. He
holds eight patents and has four more pending. He is a Fellow of IEEE, a
Fellow of SIAM and a Foreign Member of the Royal Swedish Academy of
Engineering Sciences (IVA).
\end{IEEEbiography}

\end{document}